# A Generalized Damage-Based Estimate of Crash Severity

Bob J. Scurlock, Ph.D., ACTAR
*Department of Physics, University of Florida, Gainesville, Florida*

**Introduction**

In this paper, principles of the CRASH3 damage-based collision algorithm are generalized for 3-dimensional crush profiles.

**The Vehicle Spring Model**

It is well known that a vehicle's response to impact forces can be modeled, using Hooke's law, as a linear spring whose crush is given by change-in-length, *C,* and whose resistive force is given by $F = kC$, where *k* is the "spring constant" or force required to compress the linear spring by an additional unit of length [1].

**A 2-Dimensional Vehicle Model**

We can generalize this single spring to an object extended in 2-dimensions. First, let us suppose we have a 2-dimensional vehicle model, whose sides are composed of an ensemble of linear springs. (Note reference [2] derives the relations for collisions involving two vehicles composed of linear spring ensembles.) We will label the springs within the ensemble by index $i$ = {1..N}. Using SAE conventions for the vehicle-fixed coordinate system [1], let us suppose the springs always compress parallel (antiparallel) to the *x*-axis for rear (front) impacts, and compress parallel (antiparallel) to the y-axis for driver (passenger) side impacts. Each side of the vehicle model is assigned its own subset of springs within the full ensemble. We shall neglect frictional effects associated with vehicle-to-vehicle contact in developing our model below. For a given side, let us suppose our vehicle model undergoes a collision with another object. We expect as a function of time, *t,* the ensemble of springs might undergo compression characterized by the set of changes-in-spring-lengths $\{C_i(t)\}$. The total force on our object at any instant in time, *t*, is given by:

$$F(t) = \sum_{i=1}^{N} k_i C_i(t) \qquad (1)$$

where the sum is performed over the full ensemble of *N* springs. Here $k_i$ is the spring-constant for the $i^{th}$ spring in the ensemble.

The maximum force during impact is then given by:

$$F^{Max} = \sum_{i=1}^{N} k_i C_i^{Max} \qquad (2)$$

where $C_i^{Max}$ is the maximum crush imparted to the $i^{th}$ spring during the collision. The maximum crush would generally be expected to occur at the moment of "maximum engagement" between two colliding objects. At the moment of maximum engagement, the energy absorbed compressing the $i^{th}$ spring of our object under consideration is given by the integral:

$$E_i = \int_0^{C^{Max}} dC'_i k_i C'_i = \frac{1}{2} k_i (C_i^{Max})^2 \qquad (3)$$

The total energy absorbed then is given by the sum over the full ensemble:

$$E = \sum_{i=1}^{N} E_i = \sum_{i=1}^{N} \frac{1}{2} k_i (C_i^{Max})^2 \qquad (4)$$

In the continuous limit, we can define a maximum crush profile, $C^{Max}(w)$, characterizing the crush damage as a function of location along a given side. Letting $k_i \to K = \frac{dk}{dw}$, which is the spring-stiffness per unit width along the side, our maximum spring-force becomes:

$$F^{Max} = \int_0^W dw' \cdot K C^{Max}(w') \qquad (5)$$

In differential form, the above integral can be expressed by:

$$\frac{dF^{Max}}{dw} = K C^{Max}(w) \qquad (6)$$

The energy absorbed at the moment of maximum engagement is then given by the double integral:

$$E = \int_0^W dw' \int_0^{C^{Max}} dC' K C'(w')$$
$$= \int_0^W dw' \frac{1}{2} K C^{Max}(w')^2 \qquad (7)$$

Let us now distinguish between the crush at the moment of maximum engagement, $C^{Max}$, and the permanent crush imparted to the vehicle that remains after the collision (i.e., the crush one measures at the salvage yard), $C^P$. Let's suppose the two are related by:

$$C^{Max} = C^P + \delta \qquad (8)$$

where $\delta$ can be understood as some restored length due to restitution. Re-writing equation (6) above, we have:

$$\frac{dF^{Max}}{dw} = K C^{Max}(w) = K \cdot (C^P + \delta) = K C^P + K\delta \qquad (9)$$

Let us define the "stiffness coefficients" *A* and *B* by:

$$A = K\delta \text{ and } B = K$$

It is noted that *A* (units [Force]/[Length]) is typically understood as the "pre-load" force term, characterizing the force per unit width threshold for permanent deformation. B (units [Force]/[Area]) is understood as the generalized spring stiffness value characterizing the spring resistance to compression per unit width. Using staged collision data to characterize the vehicle stiffness constant value *k*, *B* can be approximated by $B = k/W$, where *W* is the width of direct contact damage. Using this *B* along with knowledge of the no-damage closing-speed limit, *A* can be obtain. This approximation of *A* and *B* assumes a homogenous stiffness response by the contact surface; that is, that both *A* and *B* are constant as a function of *w*.

Here we note the above implies $\delta = A/B$, thereby allowing us to rewrite maximum crush as:

$$C^{Max} = C^P + \frac{A}{B} \tag{10}$$

This yields the familiar form given by the CRASH3 model [3]:

$$\frac{dF^{Max}}{dw} = A + B \cdot C^P = B \cdot \left(\frac{A}{B} + C^P\right) \tag{11}$$

Returning to our total maximum force calculation, we have:

$$F^{Max} = \int_0^W dw' \cdot (A + B \cdot C^P(w')) \tag{12}$$

and total absorbed energy at the moment of maximum engagement:

$$E = \frac{1}{2} \int_0^W dw' \cdot B \cdot \left(\frac{A}{B} + C^P(w')\right)^2 \tag{13}$$

In our discretized model, the energy absorbed by the $i^{th}$ spring can be written now as:

$$E_i = \frac{1}{2} k_i (C_i^{Max})^2 = \frac{1}{2} \Delta w_i \cdot B \cdot \left(C_i^P + \frac{A}{B}\right)^2$$
$$= \Delta w_i \cdot \left(\frac{B \cdot (C_i^P)^2}{2} + \frac{A^2}{2 \cdot B} + A \cdot C_i^P\right) \tag{14}$$

or,

$$\boldsymbol{E_i = \Delta w_i \cdot \left(\frac{B \cdot (C_i^P)^2}{2} + \frac{A^2}{2 \cdot B} + A \cdot C_i^P\right)} \tag{15}$$

where $\Delta w_i$ is the width associated with the $i^{th}$ spring. Visualizing each spring as being contained within a "cell" (or "bin") of width $\Delta w_i$, the crushed area within the $i^{th}$ cell at the moment of maximum engagement can be approximated by the product

$$Area_i = \Delta w_i \cdot \left(\frac{A}{B} + C_i^P\right)$$

The energy absorbed per unit width is then given by the usual expression:

$$\frac{E_i}{\Delta w_i} = \frac{B \cdot (C_i^P)^2}{2} + \frac{A^2}{2 \cdot B} + A \cdot C_i^P \tag{16}$$

And the total energy absorbed is given by:

$$\boldsymbol{E = \sum_{i=1}^N \Delta w_i \cdot \left(\frac{B \cdot (C_i^P)^2}{2} + \frac{A^2}{2 \cdot B} + A \cdot C_i^P\right)} \tag{17}$$

**A 3-Dimensional Vehicle Model**

Let us now generalize this formalism to 3-dimensions. Suppose we have a 3-dimensional vehicle model whose surface is composed of an ensemble of linear springs distributed in 3-dimensional space across the 2-dimensional vehicle surface. Here we continue using SAE conventions for the vehicle-fixed coordinate system whereby we suppose the springs always compress parallel (antiparallel) to the *x*-axis for rear (front) impacts, and compress parallel (antiparallel) to the *y*-axis for driver (passenger) side impacts. Again, each side of the vehicle model is assigned its own subset of springs within the full vehicle ensemble.

Letting our 3-deminsional vehicle model undergo a collision, we can express the maximum force on the vehicle by using equation (2), were the sum extends over all springs in the ensemble. The energy absorbed at maximum engagement is given by equation (4).

Going to the continuous limit, we can define a maximum crush profile along a given side by the 2-dimensional function, $C^{Max}(w,h)$, which characterizes the crush damage as a function of location in height and width along the given side. Letting $k_i \to \widetilde{K} = \frac{dk}{dwdh}$, gives the spring-stiffness per unit area within the solid object face. With this, our maximum force in the continuous limit becomes:

$$F^{Max} = \int_0^H dh' \int_0^W dw' \cdot \widetilde{K} C^{Max}(w', h') \tag{18}$$

and the absorbed energy at maximum engagement becomes:

$$E = \int_0^H dh' \int_0^W dw' \int_0^{C^{Max}} dC' \widetilde{K} C'(w', h')$$
$$= \int_0^H dh' \int_0^W dw' \frac{1}{2} \widetilde{K} C^{Max}(w', h')^2 \tag{19}$$

Again, rewriting the above in differential form, we now have:

$$\frac{dF^{Max}}{dwdh} = \widetilde{K} C^{Max}(w, h) \tag{20}$$

Using equation (8), this becomes:

$$\frac{dF^{Max}}{dwdh} = \widetilde{K} C^{Max}(w, h) = \widetilde{K} C^P + \widetilde{K} \delta \tag{21}$$

Here we will identify $\widetilde{A} = \widetilde{K}\delta$ and $\widetilde{B} = \widetilde{K}$. We again note that our generalized "A" coefficient, $\widetilde{A}$ (units [Force]/[Area]) is the "pre-load" force term characterizing the force per unit area threshold for permanent deformation. $\widetilde{B}$ (units [Force]/[Volume]) is the generalized spring stiffness value characterizing the resistance to compression per unit volume. Again, assuming a homogeneous stiffness response by the contact surface, $\widetilde{B}$ can be estimated using staged collision data. Obtaining the value of B, one can use the approximation $\widetilde{B} = B/H$, where $H$ is the height of the damage region obtained from staged collisions. Similarly for $\widetilde{A}$, we have $\widetilde{A} = A/H$. In this homogenous approximation both $\widetilde{A}$ and $\widetilde{B}$ are assumed constant as a function of $w$ and $h$.

With this we have:

$$\frac{dF^{Max}}{dwdh} = \widetilde{K} C^{Max}(w, h) = \widetilde{A} + \widetilde{B} \cdot C^P = \widetilde{B} \cdot \left(\frac{\widetilde{A}}{\widetilde{B}} + C^P\right) \tag{22}$$

In the continuous limit, the total peak force can be written as:

$$F^{Max} = \int_0^H dh' \int_0^W dw' \cdot (\widetilde{A} + \widetilde{B} \cdot C^P(w', h')) \tag{23}$$

and the total energy absorbed at maximum engagement as:

$$E = \frac{1}{2} \int_0^H dh' \int_0^W dw' \cdot \widetilde{B} \cdot \left(\frac{\widetilde{A}}{\widetilde{B}} + C^P(w', h')\right)^2 \tag{24}$$

In the discrete model, for a $i^{th}$ spring within a cell in the $(w,h)$ plane, we can write the energy absorbed at maximum engagement as:

$$\boldsymbol{E_i = \Delta w_i \Delta h_i \cdot \left(\frac{\widetilde{B} \cdot (C_i^P)^2}{2} + \frac{\widetilde{A}^2}{2 \cdot \widetilde{B}} + \widetilde{A} \cdot C_i^P\right)} \tag{25}$$

Again visualizing each linear spring as being contained within its own cell of size $\Delta w_i$ in width of and $\Delta h_i$ in the height, the crushed volume

within the $i^{th}$ cell at the moment of maximum engagement can be approximated by the product

$$Volume_i = \Delta w_i \cdot \Delta h_i \cdot \left(\frac{\tilde{A}}{\tilde{B}} + C_i^P\right) \quad (26)$$

Finally, the total energy absorbed at maximum engagement is given by the sum:

$$E = \sum_{i=1}^{N} \Delta w_i \Delta h_i \cdot \left(\frac{\tilde{B} \cdot (C_i^P)^2}{2} + \frac{\tilde{A}^2}{2 \cdot \tilde{B}} + \tilde{A} \cdot C_i^P\right) \quad (27)$$

*Bob Scurlock, Ph. D. is a Research Associate at the University of Florida, Department of Physics and works as a consultant for the accident reconstruction and legal community. He can be reached at BobScurlockPhD@gmail.com. His website can be found at: ScurlockPhD.com.*